# Holographic imaging through a scattering medium by diffuser-aided statistical averaging

MICHAEL J. PURCELL[1], MANISH KUMAR[1,*], STEPHEN C. RAND[1], VASUDEVAN LAKSHMINARAYANAN[2]

1 Optics & Photonics Laboratory, EECS Dept., University of Michigan, Ann Arbor, MI 48109-2099, USA
2 School of Optometry and Vision Science, University of Waterloo, Waterloo, Canada
*Corresponding author: manishk.iitd@gmail.com

**We introduce a practical digital holographic method capable of imaging through a diffusive or scattering medium. The method relies on statistical averaging from a rotating ground glass diffuser to negate the adverse effects caused by speckle introduced by a first, static diffuser or scattering medium. In particular, a setup based on Fourier transform holography is used to show that an image can be recovered after scattering by introducing an *additional* diffuser in the optical setup. This method is capable of recovering object information from behind a scattering layer in biomedical or military imaging applications.**

## 1. INTRODUCTION

The ability to image through a scattering or diffusive medium such as tissue or a hazy atmosphere is a goal which has attracted extensive attention from the scientific community. Inhomogeneous media which randomly scatter light cause severe degradation in the quality of transmitted images. Hence many approaches have been investigated trying to realize this sort of imaging capability.

Significant contributions have been made based on: holographic wave-front reconstruction techniques [1,2]; pulsed lasers and high-speed shutters for ultrafast gating purposes to extract the first-arriving light [3]; enhanced gating based on beams with reduced spatial or temporal coherence [4-8]; ultrafast parallel wave-front optimization and adaptive compensation [9,10]; guide star methodology [11]; ultrasonically encoded time-reversed light [12,13]; two photon nonlinear microscopy [14] and wave-front optimization [15]. These earlier methods either make use of specialized ultrafast lasers or tend to be computationally intensive. Recently, many non-invasive imaging approaches have been developed which make use of the speckle phenomena arising from the scattering layer itself [16-19]. In particular, there have been approaches based on speckle correlation which utilize the principle of memory effects in scattering from a thin layer of diffusing media [17,18] or use spatial input–output correlation to identify specific back-scattered waves from objects hidden deep inside scattering media [19]. Another approach makes use of two point intensity correlation (i.e. fourth order speckle statistics) to retrieve the complex wave-field information beyond a scattering media [20]. These correlation-based methods have been promising on many fronts as they involve one-step processes which minimize the computation required to realize the final image.

While all these methods have advantages and drawbacks, a particularly simple and straightforward holographic imaging method was proposed very recently [21]. The method makes use of the ergodicity of speckle patterns and is based on statistical averaging. It uses coherent laser beams to record the holographic information of the hidden object and gives a reconstruction with suppressed speckle contrast. However, a major drawback of this method is that during the recording of the hologram, the main scattering layer must be moving or rotating. This requirement renders the method impractical for most imaging applications.

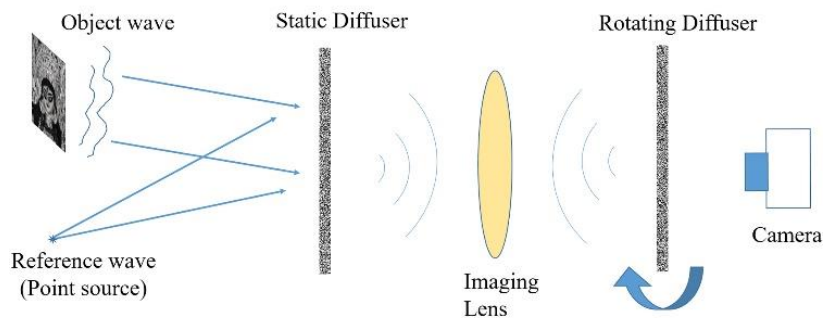

Fig.1: Schematic representation of our imaging approach.

In this paper we remove this drawback without relying on a pulsed laser as in gating based imaging or any non-linear response from the medium or target. The method also does not require motion of the main scattering layer. Instead, imaging is made possible through a static scattering surface by introducing an extra diffuser in such a way that the main diffuser is perfectly imaged on this second diffuser. By moving or rotating the second diffuser rather than the main scattering layer (see Fig. 1) statistical averaging of the speckle pattern from the object can be performed, thereby resulting in retrieval of the image even after severe scattering.

## 2. THEORY

At the input surface of the first diffuser a superposition is made of the reference and object waves, forming the hologram function

$$u(\xi,\eta) = u_{obj}(\xi,\eta) + u_{ref}(\xi,\eta). \tag{1}$$

This hologram function describes the total field just prior to the static diffuser (SLM in Fig. 1). The diffuser introduces a random phase distribution $\phi_r(\xi,\eta)$ to the incident wave-field, so the wave-field just after the SLM may be represented as

$$u_{slm} = u(\xi,\eta)e^{-i\phi_r(\xi,\eta)}. \tag{2}$$

Now, a second diffuser is introduced and the static diffuser is perfectly imaged onto it by making use of a 4-$f$ imaging system. Thus the field immediately after the second diffuser is given by

$$u_{diff} = \exp[-i\phi'_r(\xi,\eta)] \iint u(\hat{\xi},\hat{\eta}) \exp[-i\phi_r(\hat{\xi},\hat{\eta})] h_0(\xi-\hat{\xi},\eta-\hat{\eta}) \, d\hat{\xi}d\hat{\eta}, \tag{3}$$

where $\phi'_r(\xi,\eta)$ is a random phase introduced by the second diffuser and $h_0(\xi,\eta)$ is a finite spread impulse response function of the 4-$f$ imaging system. If the SLM pixel width is large enough, it leads to a relatively slow variation of the wave-field just after SLM i.e. $u(\hat{\xi},\hat{\eta}) \exp[-i\phi_r(\hat{\xi},\hat{\eta})]$ doesn't vary much within the spread of $h_0(\xi,\eta)$. Under such approximation, the impulse response function tends to a delta function and Eq. (3) reduces to a much simpler form

$$u_{diff} = u(\xi,\eta)e^{-i\phi_r(\xi,\eta)}e^{-i\phi'_r(\xi,\eta)} = u(\xi,\eta)e^{-i\phi_{tot}(\xi,\eta)}. \tag{4}$$

As we show later in experimental realization, meeting this criterion does insure better results. Now, in the case of relatively thin scattering media or reflective scattering surfaces, the object and reference waves undergo the same amount of phase distortion and scattering. Therefore, the relative phase shift between waves is maintained throughout the scattering process, with the result that information is not lost but is merely obscured by the random phase function [21]. Beyond the diffuser, the wave is recorded through imaging optics on a CCD camera. The camera captures the intensity of the field $u_{ccd}$ which is a convolution of the diffused field $u_{diff}$ and the point spread function of the imaging optics $h(\xi,\eta)$.

$$u_{ccd}(x,y) = \iint u_{diff}(\xi,\eta)h(x-\xi,y-\eta)d\xi d\eta. \tag{5}$$

The corresponding intensity is therefore expressible as

$$I_{ccd} = |u_{ccd}(x,y)|^2 = \iint \iint u(\xi_1,\eta_1)u^*(\xi_2,\eta_2)e^{-i\phi_{tot}(\xi_1,\eta_1)}e^{i\phi_{tot}(\xi_2,\eta_2)} \times h(x-\xi_1,y-\eta_1)h^*(x-\xi_2,y-\eta_2)d\xi_1 d\eta_1 d\xi_2 d\eta_2. \tag{6}$$

Because the detection optics has a finite collection angle, portions of the wave-front diffracted at large angles which are included in Eq. (6) may not be captured by the camera. Some information loss takes place due to this limitation. To reduce speckle noise, time averaging of the recorded field intensity can be performed [21,22]. This is done by rotating the second diffuser while recording the intensity with the camera. Assuming that the diffuser creates an ergodic speckle pattern which is delta correlated, we may replace time averaging with ensemble averaging [22], leading to

$$\langle e^{-i\phi_{tot}(\xi_1,\eta_1,t)}e^{i\phi_{tot}(\xi_2,\eta_2,t)}\rangle = \delta(\xi_1-\xi_2,\eta_1-\eta_2). \tag{7}$$

Then the intensity captured by the camera reduces to a convolution of the intensity of the hologram function with the detector impulse response [21,23].

$$\langle I_{ccd}\rangle = \langle |u_{ccd}(x,y)|^2\rangle = \iint |u(\xi,\eta)|^2 \times |h(x-\xi,y-\eta)|^2 d\xi d\eta. \tag{8}$$

In the equation above, the first term $|u(\xi,\eta)|^2$ is the intensity of the hologram function and the second term is the intensity impulse response of the imaging optics. Based on the convolution theorem, note that the Fourier transform of the camera intensity distribution can be expressed as a product of the Fourier transform of the hologram function (i.e. F.T.$\{|u(\xi,\eta)|^2\}$) and the Fourier transform of the impulse response function (i.e. F.T.$\{|h(x,y)|^2\}$) which is the optical transfer function (OTF) of imaging optics [23].

## 3. EXPERIMENT

If the hologram is recorded in the lens-less Fourier transform holographic arrangement, the object wave can be easily reconstructed by simply taking a Fourier transform of the hologram function where both real and virtual images are obtained on two sides of the central DC term [21]. We follow this same principle while recording the hologram in our experimental setup, shown schematically in Fig. 2. For the experiment, an image (of 'Durga') was printed on a transmissive plastic sheet (of dimension 1 inch by 1 inch) to serve as the object. A linearly-polarized, optically-pumped semiconductor laser (Coherent Sapphire) operating continuously at 488 nm served as the light source.

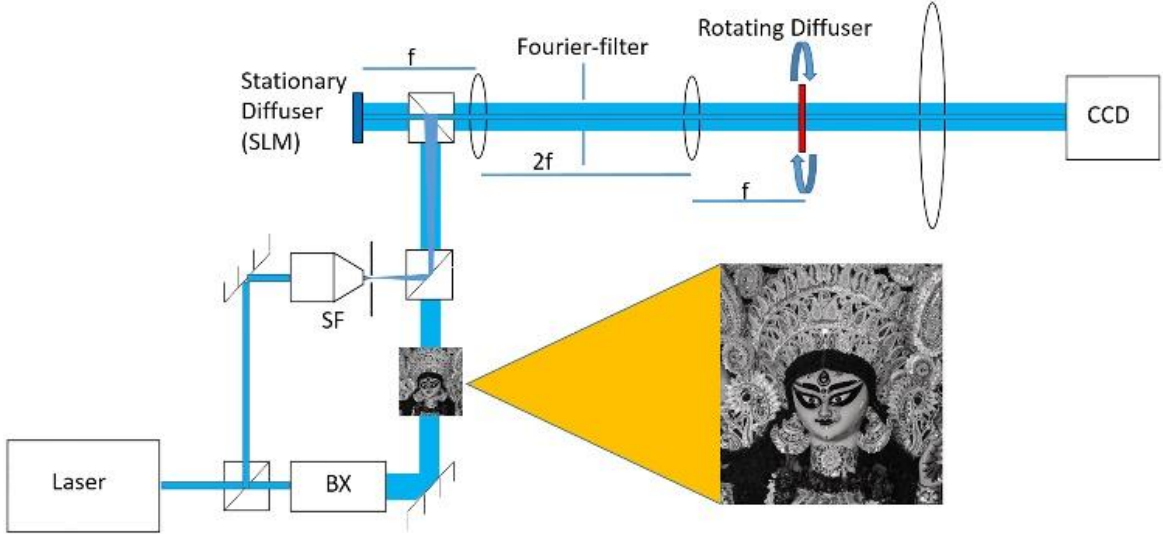

Fig. 2: Schematic of the experimental setup. SF: spatial filter, BX: Beam expander.

The light beam was split into two parts with a 50-50 non-polarizing beam splitter. One part was magnified in a beam expander and passed through the object to provide the object wave. The other part was directed through a spatial filter consisting of a 20x microscope objective and a 15 μm pinhole to provide a diverging wave-front as the reference wave. To implement the Fourier transform holographic recording approach, the pinhole (serving as the origin of the diverging reference wave) and the object were placed at equal distances from the reflective, phase-only SLM (Holoeye Pluto VIS). The reference and object waves were then combined with the help of a beam splitter to form a hologram on the SLM. The surface of the SLM was in turn imaged on an additional diffuser using a 4-$f$ imaging system (f1 = 30 mm and f2 = 12 mm). Finally, the surface of the second diffuser was imaged on a CCD camera (Andor iXon 885 EMCCD) using another dual lens based 4-$f$ imaging system. Fig. 2 shows a schematic diagram of the entire experiment with the second 4-$f$ system represented by a single lens. The SLM was used as a diffuser by applying random phases between 0 and 2π to its pixels. The Holoeye SLM has square pixels arranged in a rectangular matrix 1920 pixels wide and 1080 high over its active display area. Each pixel was 8 μm on a side. During the experiment, many adjacent pixels were grouped together to form square super-pixels. The size of super-pixels was varied from 1 pixel (8 μm) to 50 pixels (400 μm) per side. For each size, intensity data was captured on the camera and a 2-D Fourier transform was performed to reconstruct the image of the object. These measurements constituted a first set of results in which only one diffuser (the SLM) was inserted in the setup. To process background speckle, a 600-grit ground glass diffuser (Thorlabs DG20-600-MD) was placed at the image plane of the 4-$f$ configuration. This additional diffuser was mounted on a platform normal to the optical axis of the system and rotated at around two hundred rotations per minute (200 RPM) while intensities were again recorded at the camera, providing a second set of results with longer exposure times (~0.9 seconds) corresponding to the time averaged intensity calculated in Eq. (8).

Since the active area of the SLM is structurally periodic, it produces multiple orders of diffraction. Hence reflected light splits into orders that travel at different angles with respect to the optical axis in the experiment. The higher orders correspond to higher spatial frequencies and make larger angles with respect to the optical axis. Since good imaging of light from the first diffuser plane onto the second diffuser is crucial for successful image retrieval by this method, it is important that multiple diffraction orders are collected by the 4-$f$ imaging optics. To demonstrate this feature, we took advantage of pixel grouping to control the number of diffraction orders accepted by the limited numerical aperture of our experiment. Smaller sized super-pixels corresponded to higher spatial frequencies, leading to larger propagation angles with respect to the optical axis. High quality reconstruction of images scattered by small scattering centers therefore calls for imaging lenses with large numerical aperture. The highest transverse spatial frequencies are generated from the smallest groupings of diffracting structures i.e. individual pixels of the SLM. Since these pixels were 8 μm on each side, the highest spatial frequency (for first order diffraction) produced by the SLM was determined by this length.

The first lens in the 4-$f$ imaging system performs a Fourier transform of the intensity distribution coming from the SLM. The Fourier plane lies at the front focal plane of the lens, whereas the SLM is positioned at the back focal plane. Since SLM represents a sampled signal, it generates multiple copies of the band-limited signal in the Fourier plane and we need to select only the fundamental order of the bandlimited signal. For this purpose, a spatial filter was placed at this Fourier plane to exclude all copies of the sampled phase function displayed on SLM. Thus all the spatial frequencies higher than 0.5 times ±1 diffraction order for 8 μm periodicity had to be blocked. The filter was prepared by removing a square cut-out from an opaque sheet of paper. Its size was easily calculated from the following relations:

$$d\sin\theta = m\lambda, \tag{8}$$

and

$$\tan\theta = y/f, \tag{9}$$

where $d$ = periodicity of the grating on the SLM, $\theta$ is the angle made by the $m^{th}$ order with the optical axis, $\lambda$ is the wavelength and $y$ is the distance of the diffraction spot from the center of the Fourier plane. Using the above equations we find $y$=18.33 mm for the side length of the square aperture of the Fourier filter where $m$ = 1, $d$ = 8 µm and $f$ = 300 mm. Since the SLM is pixelated, the field reflected from it represents sampled data, and any function displayed on the SLM gets diffracted into multiple orders. Without pixel grouping/binning, only the zero order diffraction term was allowed to pass through the Fourier filter. For 2×2 pixel binning, up to first order diffraction terms (0, ±1) were allowed and so on. The fidelity with which a phase function can be imaged to the conjugate plane is determined by the number of diffraction orders allowed through the system since adequate sampling of the discrete signal requires the inclusion of about eight diffraction orders for good reconstruction. In our experimental setup, system performance should begin to degrade rapidly for segment sizes smaller than 16×16 pixels because this corresponds to eight diffraction orders. In the next section it is demonstrated that performance does indeed degrade for groupings of less than 15×15 pixels (16×16 pixel binning was not used).

When the SLM was addressed with random phases to act as a diffuser, the data reaching the CCD was overlaid with a speckle pattern which obscured the reconstructed image. The extent of obscuration was inversely proportional to the number of pixels grouped together in blocks on the SLM. Placing the rotating diffuser at the image plane of the 4-$f$ system was intended to counteract the effects of scattering from the SLM by performing a simple statistical averaging. The importance of precisely imaging the SLM face onto the rotating diffuser cannot be overstated. For this reason, the positioning of the rotating diffuser was critical. Experimentally it was found that small offsets in diffuser position from the imaging plane caused serious degradation of system performance due to fall in the OTF in Eq. (8).

## 4. RESULTS AND DISCUSSION

To eliminate the limitation due to numerical aperture and obtain the best image possible in our experimental setup, all the SLM pixels were binned together as one super-pixel. In this case the SLM was a blank screen with uniform phase that behaved like a plane mirror (except for the thin circuit lines around each pixel). With such a blank image on the SLM, we took a long exposure of 0.9 seconds through the rotating diffuser with the camera. By making use of a fast Fourier transform (FFT) algorithm in MATLAB we then reconstructed the image. Fig. 3 shows the result, which suitable as a point of reference in the present work. This image may also serve as a point of comparison with the results in ref. 21 since the only diffuser in the system is in motion. Note that the cropped region in Fig. 3 is used throughout this paper in evaluating other results.

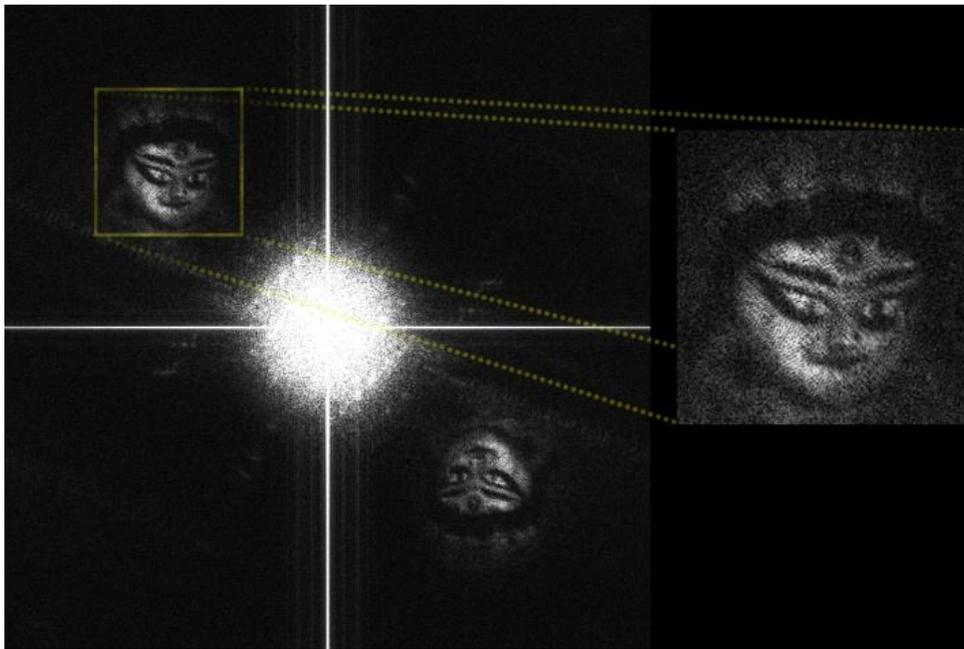

Fig. 3: The reconstructed image for the case of blank SLM with rotating diffuser. Inset shows the magnified view of the recovered image which serves as a reference for our analysis.

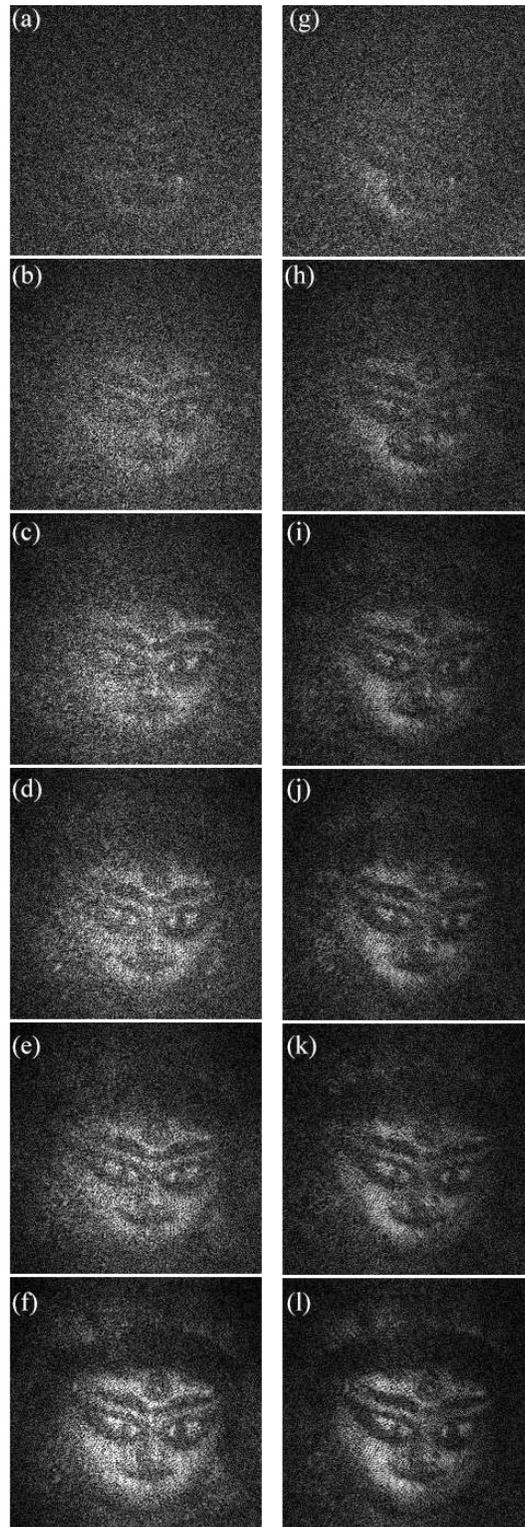

Fig. 4: The recovered image comparison for varied pixel binning sizes of 5×5, 10×10, 15×15, 20×20, 25×25 and 50×50 (top to bottom) for two cases: without rotating diffuser in (a)-(f) and with rotating diffuser (g)-(l) in the setup.

Next, we made two sets of measurements with various groupings of SLM pixels. The first set comprised results obtained without the rotating diffuser in place. That is, the static diffuser (SLM) was in place but there was no possibility of statistical averaging. Results were recorded with bin-sizes of 5×5, 10×10, 15×15, 20×20, 25×25 and 50×50 pixels and cropped reconstructed images are shown in the left column of Fig. 4. A second set of results was obtained by placing a rotating diffuser at the image plane of the first 4-$f$ setup. On the basis of our earlier analysis, this should improve the image quality. The results are shown in the right column in Fig. 4, opposite the unaveraged images for the same super-pixel sizes. A visual inspection is sufficient to conclude that, for a given bin size, inserting a rotating diffuser gives an improved result. However the quality of these results can also be evaluated more quantitatively using a numerical procedure.

Table 1: Comparison of SSIM data for assessing improvement in image recovery

| Binning size | Without rotating diffuser (SSIM) | With rotating diffuser (SSIM) | Percentage improvement in SSIM |
|---|---|---|---|
| 5×5 | 0.0816 | 0.0894 | 9.6 |
| 10×10 | 0.0962 | 0.1912 | 98.8 |
| 15×15 | 0.1166 | 0.2618 | 124.5 |
| 20×20 | 0.1372 | 0.3103 | 126.2 |
| 25×25 | 0.1680 | 0.3108 | 85.0 |
| 50×50 | 0.1904 | 0.3932 | 106.5 |

The structural similarity index [24] was chosen to evaluate image quality. In this method each result is compared to a reference image, such as that in Fig. 3. We made use of MATLAB to compute the structural similarity index (SSIM) [25] and have listed the SSIM values for all images in Table 1. In agreement with the visual impression, the SSIM data confirms that there is a significant improvement in image quality for binning of more than 10×10 pixels. This is a striking confirmation of the theory. Based on the table, it may be noted further that when more than 10×10 pixels are binned together, the quality of reconstruction does not improve much in the case where a rotating diffuser is not employed. This conclusion is not so obvious, since the series of images presented in Fig. 4 appear to show substantial improvement to the eye in terms of quality of recovered image. Upon closer inspection it may be verified that it is only larger features of the image which show improved recovery for pixel groupings above ten. Since the SSIM index evaluates similarities down to the individual pixel level, it provides a more globally accurate assessment of image quality than is possible with the unaided eye. One may be interested in knowing why the recovery of images even with rotating diffuser worsens with reduction in the pixel binning size. This has to do with the impulse response function $h_0(\xi,\eta)$ of the imaging optics between SLM and rotating diffuser as discussed around Eq. (3). As the binning size of SLM is reduced, the variation of wave-field beyond this SLM (random static diffuser) becomes fast enough in comparison to the impulse response function $h_0(\xi,\eta)$ leading to its approximation as a delta function no more valid. This in turn makes the assumption of field just after second diffuser being same as hologram function multiplied by a random phase only function (as expressed in Eq. (4)) invalid. Thus the intensity recorded by camera can no more represent the hologram accurately tending to no recovery of the object information. For this very reason the recovery of object information became negligible for binning size of lesser than 5×5 pixels. While the set of experiments helps in understanding and verifying these implications, it is pretty straight forward to understand that the requirement of imaging through a smaller binning size would require a narrower impulse response function $h_0(\xi,\eta)$. This is achievable by using a higher numerical aperture and better aberration compensated optics.

In order to add more value to our conclusion that introduction of additional moving diffuser helps in recovery of smaller features from the object, we replaced the object with a USAF resolution target and repeated same experiments. The SLM binning size was kept at 5×5 pixels. Whole setup was kept same and the Fourier holograms were captured in camera with exposure time of 0.23 second. Here, a relatively lower exposure time was possible due to the choice of ND filters in the setup which ensured the availability of sufficient light intensity even at lower exposure time. Once again the images were reconstructed by a single fast Fourier transform (FFT) operation on the recorded images. Fig. 5 shows the reconstructed images of this experiment. It is clearly seen that for a given bin size, introduction of rotating diffuser leads to more improved imaging and recovery of smaller feature size.

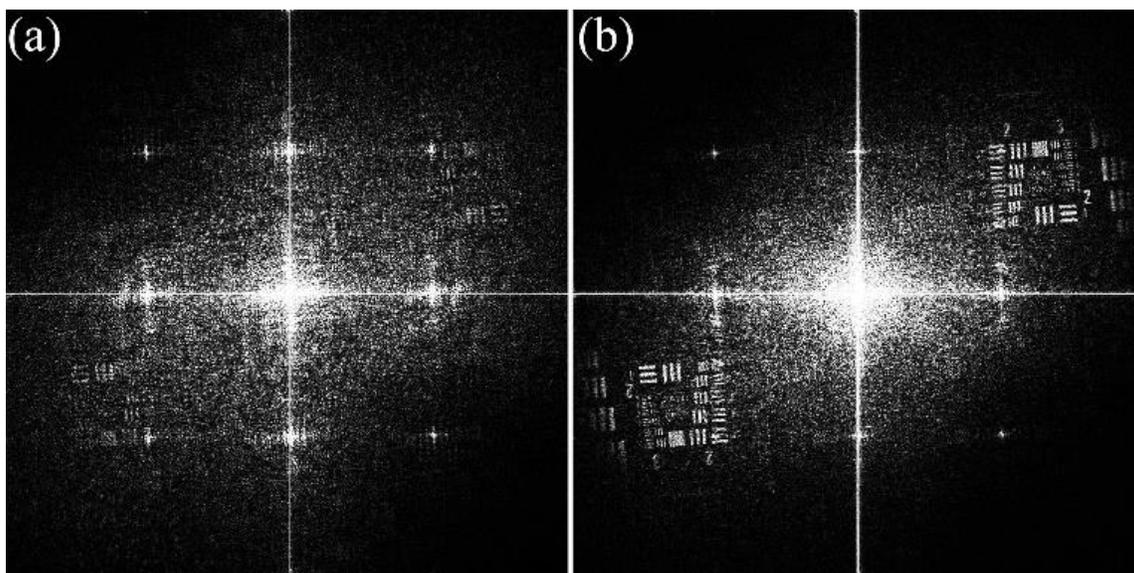

Fig. 5: Image reconstruction of USAF resolution target for 5×5 pixel binning in SLM. In (a) we have the reconstruction for the case of no rotating diffuser while in (b) the result with introduction of a rotating diffuser in presented.

In addition to the experiments described above, we attempted image recovery after sandwiching the object between static diffusers. This situation closely resembles that of an object embedded within a scattering medium where plane wave illumination is not possible. Fig. 6 shows the schematic representation of this setup. Another 600-grit ground glass diffuser (Thorlabs DG20-600-MD) was introduced in such a way that the object lies between static diffuser 1 and the SLM which again serves as a reconfigurable static diffuser 2. The rest of the setup was unaltered and the detection and image processing remained the same. Images were again recorded for different pixel bin sizes on the SLM. A trend similar to that obtained in the previous case was found. Fig. 7 shows the result associated with a bin-size of 15×15 pixels. The SSIM index evaluation of recovered images in this case gave values of 0.0260 without the rotating diffuser and 0.0345 with the rotating diffuser. It is clear that introduction of a rotating diffuser again improves the results through statistical averaging although overall image quality noticeably worsens when compared with recovered images from experiments with a single static diffuser (listed in Table 1). This is to be expected, since in the case of an object sandwiched between diffusers the illumination is by a speckle field. Hence it is illuminated non-uniformly. However, this is not an insurmountable or fundamental issue, because modulation of the incident beam could be used to make the object illumination more uniform whenever data collection is time-averaged. While all the improvements in imaging reported here are purely due to the advantages of the optical method itself, it is obvious that additional improvements in the technique could result from further computer processing.

The main limitation of the technique reported here arises from the challenge of precisely imaging the static diffuser onto the rotating diffuser. In practice, this limits the binning size of pixels in our experiments. We were able to work only with pixel groupings down to about ten pixels while maintaining acceptable image recovery. The size of a super-pixel containing ten pixels on our SLM corresponds to scattering centers with approximate diameters of 80 μm. As a limiting grain size of scattering centers, this is much bigger than what is representative of standard diffusers or tissue phantoms. On the other hand this is a limit which can readily be overcome to allow one to image through a regular diffuser. For example the imaging optics in our setup can be replaced with high numerical aperture lenses/microscope objectives to make the technique practical in scattering media with smaller effective scattering centers.

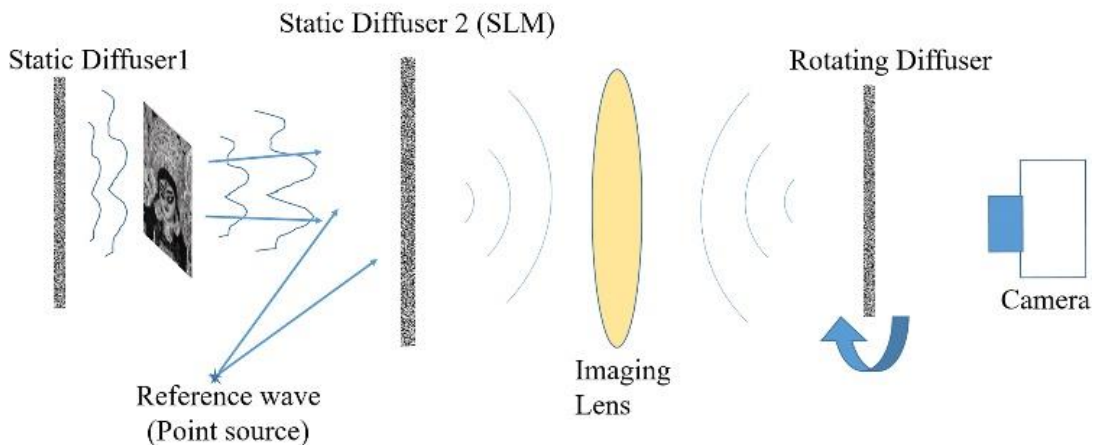

Fig. 6: Schematic of the experiment for imaging an object sandwiched between two static diffusers.

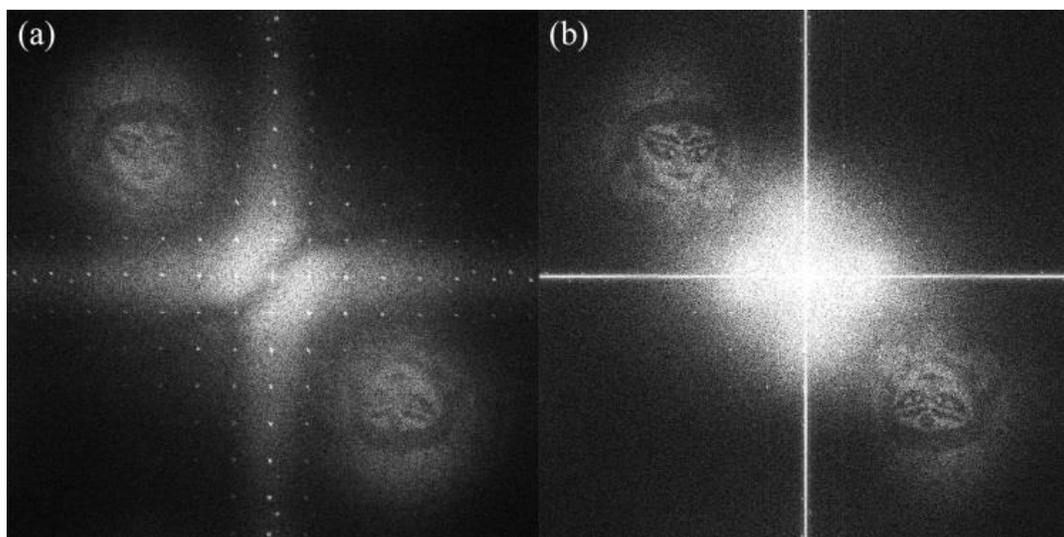

Fig. 7: The recovered image comparison for 15×15 pixel binning size in case of object sandwiched between two-diffuser configurations. Two cases: without additional rotating diffuser in (a) and with additional rotating diffuser (b).

Only the imaging system situated between the static and rotating diffuser needs to be perfect. The camera itself does not need to incorporate high numerical aperture optics. Moreover, it is possible to combine our approach with one proposed by Goodman [1] to improve it further. As shown by Goodman, for Fourier transform based holographic recording, as the distance to the sensor (photographic film in his case) is reduced, it is possible to record the hologram perfectly. This effect could be adapted in our approach to overcome the limit on optical imaging outlined above. By placing the rotating diffuser very close to the static diffuser and imaging the rotating surface on a camera with low numerical aperture optics, the need for high numerical aperture imaging could be eliminated altogether. In this scenario additional unanticipated applications would even be enabled. For example, the use of a sacrificial rotating diffuser might make it possible to image through corrosive or adverse environments where the placement of a sensor next to a containment wall might not be feasible. Unlike several recent imaging methods which utilize high order correlations or memory effects [17,18], the general approach described here is capable of imaging macroscopic objects. Other advantages of this approach include its speed, which is limited only by the speed of rotation of the diffuser and amount of available light. In our experiments, the diffuser was rotated at 200 RPM and no direct effort was made to maximize the light throughput. This resulted in an exposure time of approximately 0.23 second (for the results with USAF resolution target in Fig. (5)). At higher rotation speeds and higher light intensity, the required exposure time can be greatly reduced. It is possible to extend the method for non-holographic imaging as well where one uses the object wave alone. But in such a case the phase information of the object wave is lost and one would need to make use of Fienup type iterative algorithm to reconstruct the object which is time consuming. In our holographic imaging approach, since the reconstruction process involves only a single FFT operation on the acquired data, imaging at video refresh rates should be possible. The holographic approach opens up the possibility of imaging phase part of objects as well along with its amplitude. Finally, our method does not has any wavelength-dependence, so it could be implemented at microwave, radio or acoustic frequencies, enabling new applications in many fields.

**Funding Information**: We wish to acknowledge equipment funding from ONR DURIP (N00014-12-1-0933).

**Acknowledgment**: M. Purcell gratefully acknowledges Applied Physics Fellowship.

## References

1. J. W. Goodman, W. H. Huntley Jr., D. W. Jackson and M. Lehmann, "Wavefront-reconstruction imaging through random media," Applied Physics Letters **8**, 311-313 (1966).
2. H. Kogelnik and K. S. Pennington. "Holographic imaging through a random medium," JOSA **58**, 273-274 (1968).
3. M. A. Duguay, and A. T. Mattick. "Ultrahigh speed photography of picosecond light pulses and echoes," Applied optics **10**, 2162-2170 (1971).
4. N. Abramson, "Light-in-flight recording by holography," Optics Letters **3**, 121-123 (1978).
5. N. H. Abramson and K. G. Spears. "Single pulse light-in-flight recording by holography," Applied optics **28**, 1834-1841 (1989).
6. E. Leith, H. Chen, Y. Chen, D. Dilworth, J. Lopez, R. Masri, J. Rudd, and J. Valdmanis, "Electronic holography and speckle methods for imaging through tissue using femtosecond gated pulses," Applied optics **30**, 4204-4210 (1991).
7. E. N. Leith, D. Dilworth, C. Chen, H. Chen, Y. Chen, J. Lopez, and P. C. Sun, "Imaging through scattering media using spatial incoherence techniques," Optics letters **16**, 1820-1822 (1991).
8. E. Leith, C. Chen, H. Chen, Y. Chen, D. Dilworth, J. Lopez, J. Rudd, P. C. Sun, J. Valdmanis, and G. Vossler, "Imaging through scattering media with holography," JOSA A **9**, 1148-1153 (1992).
9. M. Cui, "Parallel wavefront optimization method for focusing light through random scattering media," Optics letters **36**, 870-872 (2011).
10. J. Tang, R. N. Germain, and M. Cui. "Superpenetration optical microscopy by iterative multiphoton adaptive compensation technique," Proceedings of the National Academy of Sciences **109**, 8434-8439 (2012).
11. D. B. Conkey, A. N. Brown, A. M. Caravaca-Aguirre, and R. Piestun, "Genetic algorithm optimization for focusing through turbid media in noisy environments," Optics express **20**, 4840-4849 (2012).
12. Y. M. Wang, B. Judkewitz, C. A. DiMarzio, and C. Yang, "Deep-tissue focal fluorescence imaging with digitally time-reversed ultrasound-encoded light," Nature communications **3**, 928 (2012).
13. Y. Liu, P. Lai, C. Ma, X. Xu, A. A. Grabar, and L. V. Wang. "Optical focusing deep inside dynamic scattering media with near-infrared time-reversed ultrasonically encoded (TRUE) light," Nature communications **6**, 5904 (2015).
14. F. Helmchen, and W. Denk, "Deep tissue two-photon microscopy," Nature methods **2**, 932-940 (2005).
15. A. P. Mosk, A. Lagendijk, G. Lerosey, and M. Fink, "Controlling waves in space and time for imaging and focusing in complex media," Nature photonics **6**, 283-292 (2012).
16. I. Freund, "Looking through walls and around corners," Physica A: Statistical Mechanics and its Applications **168**, 49-65 (1990).
17. J. Bertolotti, E. G. van Putten, C. Blum, A. Lagendijk, W. L. Vos, and A. P. Mosk, "Non-invasive imaging through opaque scattering layers," Nature **491**, 232-234 (2012).
18. O. Katz, P. Heidmann, M. Fink, and S. Gigan. "Non-invasive single-shot imaging through scattering layers and around corners via speckle correlations," Nature Photonics **8**, 784-790 (2014).
19. S. Kang, S. Jeong, W. Choi, H. Ko, T. D. Yang, J. H. Joo, J. S. Lee, Y. S. Lim, Q. H. Park, and W. Choi, "Imaging deep within a scattering medium using collective accumulation of single-scattered waves," Nature Photonics **9**, 253-258 (2015).
20. R. K. Singh, R. V. Vinu, and A. Sharma, "Recovery of complex valued objects from two-point intensity correlation measurement," Applied Physics Letters **104**, 111108 (2014).
21. A. K. Singh, D. N. Naik, G. Pedrini, M. Takeda, and W. Osten. "Looking through a diffuser and around an opaque surface: A holographic approach," Optics express **22**, 7694-7701 (2014).
22. J. W. Goodman, Statistical Optics (John Wiley, 1985).
23. J. W. Goodman, Introduction to Fourier optics (Roberts and Company Publishers, 2005).
24. Z. Wang, A. C. Bovik, H. R. Sheikh, and E. P. Simoncelli, "Image quality assessment: from error visibility to structural similarity," Image Processing, IEEE Transactions on **13**, 600-612 (2004).
25. http://www.mathworks.com/